\documentclass{aa520}
\usepackage{graphicx}
\usepackage{natbib}
\bibpunct{(}{)}{;}{a}{}{,}

\begin{document}
   \title{CVcat: an interactive database on cataclysmic variables}

   \author{J. Kube\inst{1,}\thanks{\emph{Present affiliation:
          Alfred-Wegener-Institute for Polar- and Marine Research,
          Telegrafenberg A43, D-14473 Potsdam, Germany; Koldewey-Station, N-9173 Ny-\AA\/lesund, Norway}}
          \and	
	      B. T. G\"ansicke\inst{1,2}
          \and
          F. Euchner\inst{1}
          \and
          B. Hoffmann\inst{1}}

   \institute{Universit\"ats-Sternwarte G\"ottingen, Geismar Landstra\ss\/e 11, D-37083
G\"ottingen, Germany
         \and
Department of Physics and Astronomy, University of Southampton,
Hampshire, Southampton SO17 1BJ, UK
             }

   \date{Received 29 November 2002 / Accepted 12 March 2003}

   \abstract{ CVcat is a database that contains published data on
cataclysmic variables and related objects. Unlike in the
existing online sources, the users are allowed to add data to the
catalogue. The concept of an ``open catalogue'' approach is reviewed
together with the experience from one year of public usage of
CVcat. New concepts to be included in the upcoming AstroCat framework
and the next CVcat implementation are presented. CVcat can be found at
\texttt{http://www.cvcat.org}.   }

   \maketitle
%

\section{Introduction}
CVcat is an interactive database or ``online catalogue'' that offers a number
of features so far unknown to scientific catalogues. It was developed as a
tool for the research community working on cataclysmic variables (CVs), a
class of close interacting binaries, and as a case study for some of the
concepts to be used in the development of a general catalogue software,
AstroCat.
CVcat can be accessed online at \texttt{http://www.cvcat.org}. It was first
presented to the public in August 2001 during a CV conference held in
G\"ottingen \citep{KGH}. Since then, the number of users of CVcat has
increased to more than one hundred, the daily average of requests is around
fifty (total number of delivered pages).

\section{The concept}

CVcat has been developed in order to overcome major conceptual shortcomings of
existing CV catalogues: \citet{Ritter} include only systems with known
orbital period, which limits their catalogue to $\approx 1/3$ of all CVs and
related objects. \citet{Downes} do list all known CVs, but their catalogue
provides only very limited information on each individual system, i.e., the
only binary parameter included is the orbital period. Our aim was to develop
an online data base that combines the information of the existing catalogues
\emph{and} allows the users to actively contribute to the content of the data
base, implementing a first version of an ``open catalogue''.

CVcat differs from other CV catalogues and other astronomical databases in the
concept of the data input. So far, the majority of astronomical catalogues
have been compiled by \emph{relatively small editorial teams} consisting of
scientists knowledgeable in the fields covered by the catalogues
\citep[e.g.][]{Downes,Ritter,mccook+sion99-1,liuetal01-1}. These catalogues
typically contain more or less detailed information on a specific class of
astronomical objects. Updates are published, if at all, only on a very
irregular basis. The catalogues contain just \emph{one value} for each listed
property (e.g. distance, orbital period) of a given object. While
this is helpful for non-specialist users to obtain a quick overview of
the properties of an individual object, or of the statistical
properties of a given group of objects, the more expert user will
certainly benefit if different and possibly competing values for a
given parameter are referenced in such catalogues. This is particularly useful if the information has been 
obtained by different methods.

Some of the aforementioned catalogues moved from ``classical'' printed publication to online web-based
publication, which allows shorter update cycles [e.g. the ``living edition''
of the \citet{Downes} CV catalogue, \citet{LivingDownes}], however, the
overall concepts remained unchanged. In addition to these specialized
catalogues for a specific object class there exists huge data bases like SIMBAD
\citep{SIMBAD}, which provide very basic properties for an enormously large
number of objects. However, due to the very global coverage of astronomical
objects, data contained in SIMBAD are prone to be incomplete and/or
inaccurate.

The concept of an ``open catalogue'' implemented in CVcat permits \emph{every
registered user} to add data to the catalogue, which is \emph{instantly}
visible to all other users. The quality control is performed by an editorial
team (Sect.\,2.1), which may alter or remove erroneous data.  For every
property of an object, an \emph{arbitrary number of values} can be stored
(e.g. several published values for the distance). CVcat returns one of these values as the ``best
available'' value, selected as such by the editors. However, as such
a selection process often involves some subtle subjective view of the
editor, the more expert user may decide to inspect the original sources for
the competing values, and, thereafter, decide based on his/her
experience which value is best suited for a given purpose.

\begin{figure}
\begin{center}
\resizebox{7cm}{!}{\includegraphics{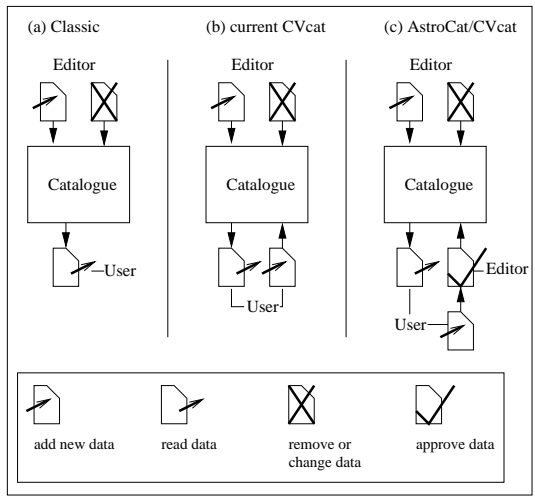}}
\caption{Classical, current, and future concepts for adding and editing
catalogue data in comparison.}
\label{f:concept}
\end{center}
\end{figure}

A sketch of the different concepts of data input and validation is given in
Fig.~\ref{f:concept} (a) for classical catalogues and (b) for the CVcat
concept. (c) introduces the concept that will be implemented in the next
release of CVcat, which is adressed in Sect.~\ref{s:prospects}. In the future
version, newly-added data will be instantly visible as it is now
[Fig.~\ref{f:concept}, (b)], but will be tagged as ``unapproved'' until an
editor cross-checks the data. In the current implementation (b), the user
cannot see if a database entry has been approved by an editor.

\subsection{Distributed editorial team}
The editorial team of CVcat is recruited from the CV research community, and
consists (ideally) of one expert per CV subclass. Since these editors are
typically familiar with the publications on their ``favourite'' objects
anyway, the amount of work to cross-check newly entered data is lowered. We
estimate that for a typical subclass of CVs, say, polars, the time to be spent on editorial duties in CVcat is of the order of two hours per week
or less.  Each editor has the privilege to remove or change erroneous data for
that person's object class only. For data on all other object classes, the editor is a
non-privileged user who may add new data and browse the
catalogue. Note that every user may add data, in contrast to,
e.g., SIMBAD, where only the editorial team can directly modify the catalogue
content.

\subsection{Database content}
In contrast to the existing CV catalogues, CVcat is designed to contain a
great variety of information for each object. Currently, the following object
properties can be included, with the number in brackets being the number of
entries on September 17, 2002 (note that because CVcat allows multiple entries
for a given property of an object these numbers do not represent the number of
unique objects for which a particular property is known):

High state magnitude                            (922),
low state magnitude                             (794),
optical spectrum exists                         (665),
orbital period                                  (509),
distance                                        (221),
general magnitude                               (209),
primary mass                                    (156),
inclination                                     (153),
secondary mass                                  (142),
superhump period                                (128),
secondary spectral type                         (121),
mass ratio                                      (93),
Doppler tomogram exists                         (80),
primary temperature                             (76),
optical light curve exists                      (61),
hydrogen column density                         (58),
orbital ephemeris                               (55),
eclipsing                                       (48),
spin period                                     (36),
primary radius                                  (33),
secondary radius                                (29),
uv data exists                                  (22),
general magnetic field strength of primary star (18),
field strength of primary magnetic pole         (17),
x-ray data exists                               (14),
eclipse map exists                              (11),
secondary temperature                           (8),
colatitude of primary magnetic pole             (5),
field strength of secondary magnetic pole       (3),
spin ephemeris                                  (3),
azimuth of primary magnetic pole               (2),
magnitude in eclipse                            (2),
azimuth of secondary magnetic pole             (1).

Every value stored in CVcat is linked to its original publication, either in
the NASA ADS using the ADS bibcode of the paper \citep{ADS}, to the
astro-ph/arXiv e-print archive, or to the VSNET messages
\citep{VSNET}. Besides references to the publications from which the data
entries contained in CVcat are taken, a list of articles with general
information on a given object can be stored in CVcat.

We allow inclusion of data from astro-ph, which is not necessarily
identical to the data
published in the final refereed version (or which may in some cases
never make it through the refereeing stage). Most of the astro-ph data,
however, is promoted to refereed information at some point. It is a typical
task of the editors to track such updates and to conduct the appropriate
changes to the database, i.e. changing the source from an astro-ph to an ADS
bibcode.

\subsection{Searching the database}

\begin{figure}
\resizebox{\hsize}{!}{\includegraphics{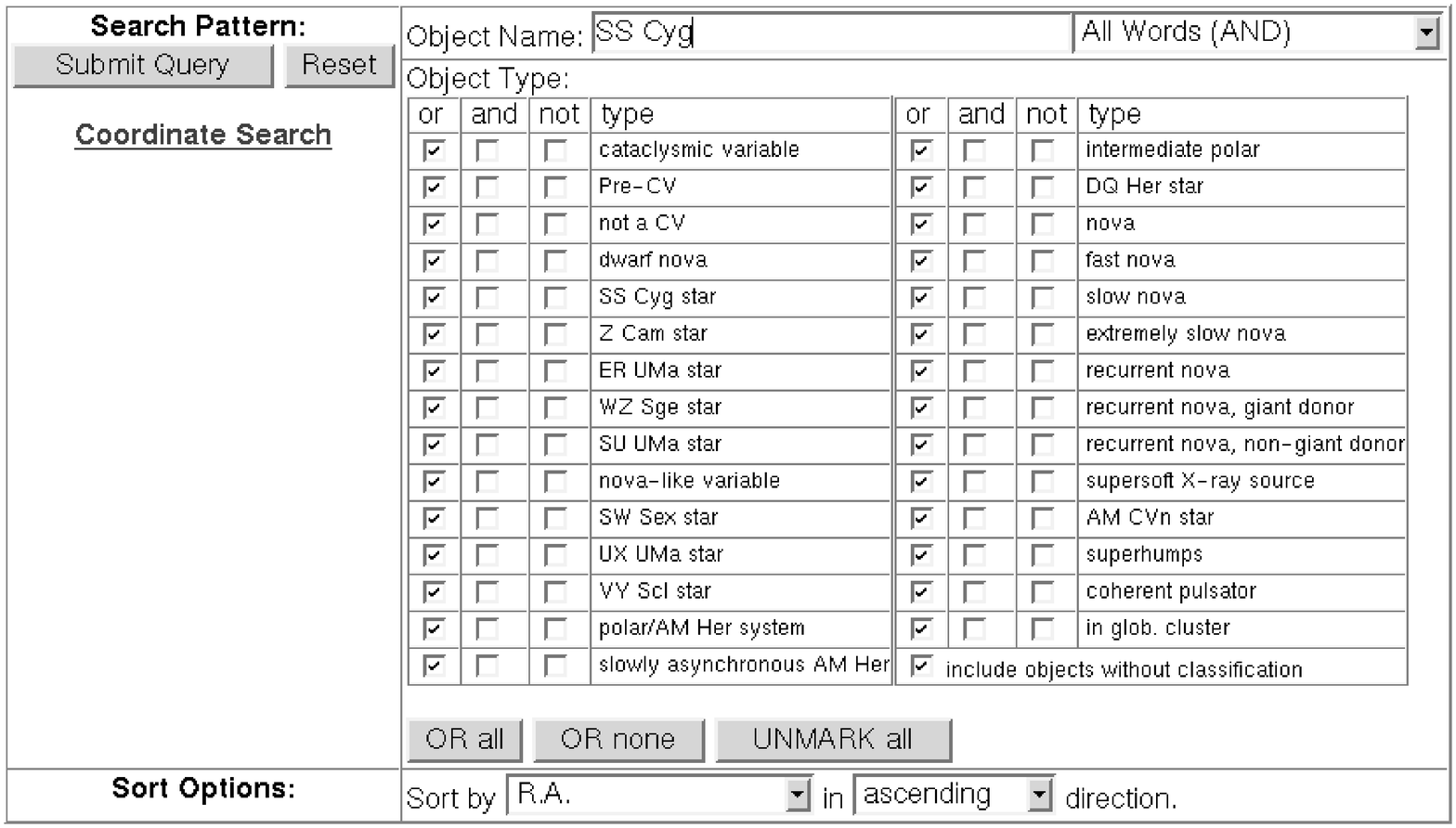}}
\caption{Search form in CVcat: A large fraction of the search form focuses on
the detailed selection of the object class.}
\label{f:search}
\end{figure}

Data retrieval from CVcat works in two ways: (i) the user obtains all
available data for a specific object, which can be found using its object
type, its coordinates, or one of its names, (ii) the user creates a table
containing selected properties for a list of objects. The latter method allows the user
to create data tables suitable for easy graphics generation as well as
ready-to-publish \LaTeX-tables.

\begin{figure}
\resizebox{\hsize}{!}{\includegraphics{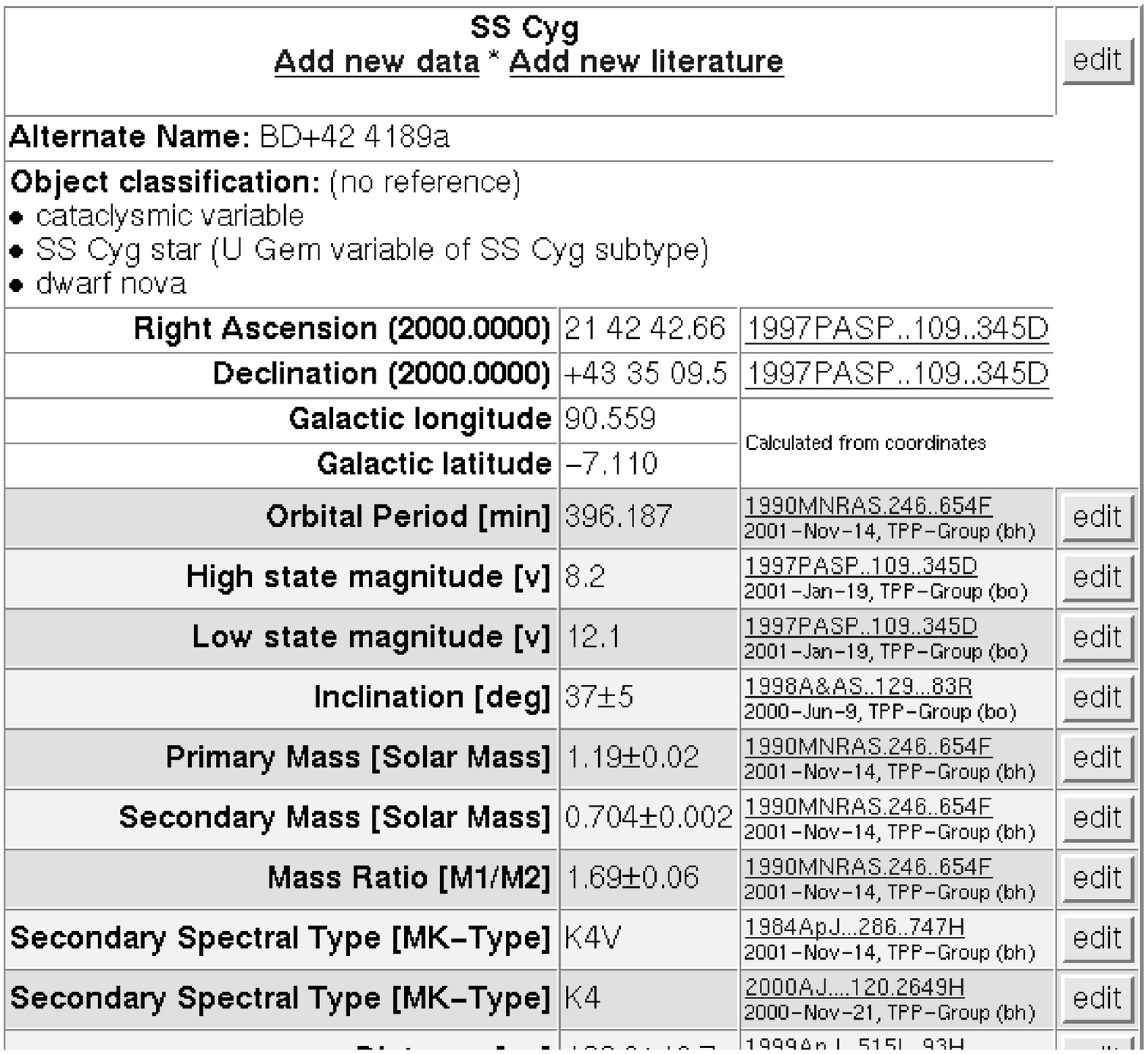}}
\caption{Search result in CVcat: Some of the data stored for SS Cygni}
\label{f:result}
\end{figure}

Searching in CVcat is organized as a two-step process. In the first step, the
user can enter the search pattern, which can be the name or a substring of the
name, a set of object classes, and the coordinate range of the object
(Fig.~\ref{f:search}). The object class is selected using a grid of logical
operators (``and'', ``or'', and ``not''). It is possible, e.g., to look for
dwarf novae, which have also been observed as novae: choose ``and DN'' and
``and Nova''. Another example is to look for nova-likes which do not show the
SW\,Sex phenomenon: the corresponding selection would be ``or NL'' and ``not
SW''.

After submitting this search request a list with all objects matching the
search criteria is returned (not shown). In this list, all object names are
hyperlinks to the page showing all results for the specific object
(Fig.~\ref{f:result}). Alternatively, a set of objects can be chosen from this
list to generate a user-configurable list with certain properties of the
objects.  This list may be adjusted in a way that objects without a published
value for a certain property are not included. By iterative calls of the list
generator it is possible to distill a table containing e.g. all objects with
known masses and periods.

An example of the results of such a list creation process is given in the
following section.

\subsection{Example: Orbital periods and donor masses}
\begin{figure}
\resizebox{\hsize}{!}{\rotatebox{270}{\includegraphics{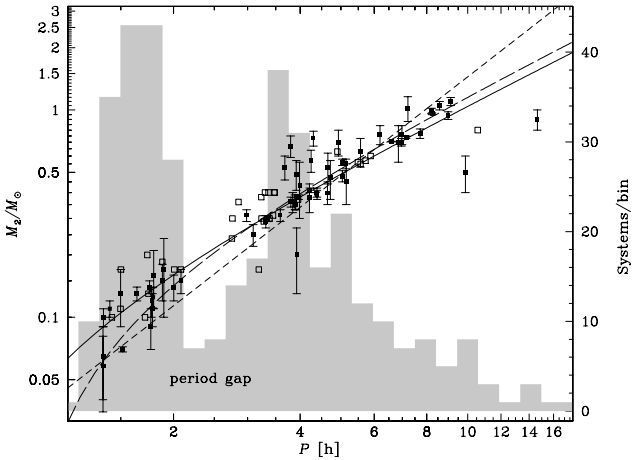}}}
\caption[Secondary star masses and period histogram]{Secondary mass plotted
against orbital period (points) and period histogram (gray bars) for all CVs with
known orbital period and secondary mass. The full line shows the relation from
Eq. (\ref{e:introcv.m2overp}), the short dashed line from
Eq. (\ref{e:introcv.m2overp_a}), the long dashed line from
Eq. (\ref{e:introcv.m2overp_b}). Full boxes and error bars are used for
systems with estimated errors in $M_2$, open boxes for secondary star masses
with an unknown error range. The given $(M_2,P)$ data pairs represent $\approx
10\%$ of all known CVs, the histogram contains $\approx 39\%$ of all CVs known
to CVcat.}
\label{f:introcv.m2overp}
\end{figure}

Using the data contained in the CVcat database, we have plotted the secondary
star masses as a function of the orbital period, Fig. \ref{f:introcv.m2overp},
for all CVs. A linear trend with
\begin{equation}
\frac{M_2}{M_{\sun}}\approx 0.11 P/{\rm h}-0.06
\label{e:introcv.m2overp}
\end{equation}
is clearly visible. This is predicted by theoretical considerations, where the
reasoning is roughly this \citep{Frank92}: Using an approximation for the
Roche geometry, valid for $1.3\la q\la 10$, and Kepler's law, one
finds that the mean density of a Roche lobe filling star is a function of only
the orbital period. With the know\-led\-ge of the lower main-sequence $M/R$
relation \citep{Kippenhahn90,BCA98}, one then finds
\begin{equation}
\frac{M_2}{M_{\sun}}\approx 0.11 P/{\rm h}
\label{e:introcv.m2overptheory}
\end{equation}

A more detailed analysis of secondary star masses leads to slightly different
period-mass-relations \citep{1998MNRAS.301..767S}:
\begin{eqnarray}
\frac{M_2}{M_{\sun}}&=&(0.038\pm0.003)(P/{\rm h})^{(1.58\pm0.09)}\quad\mbox{or}
\label{e:introcv.m2overp_a}\\
\frac{M_2}{M_{\sun}}&=&(0.126\pm0.011)P/{\rm h}-(0.11\pm0.04)
\label{e:introcv.m2overp_b}
\end{eqnarray}
The data currently available in CVcat are consistent with the results from
Smith \& Dhillon (Fig.~\ref{f:introcv.m2overp}). Note that some of the
published secondary star masses may be derived from the orbital period using
some theoretical models, hence artificially stabilizing the fit close to the
theoretical predictions.

\section{Usage statistics}

The log file analysis of CVcat usage over a one year period shows that
CVcat has a stable user community which is still slowly increasing. The typical usage of
CVcat is to request all available data on a specific object, which is normally
queried by its name. List generation of objects selected by their class ranks
second in the usage of the database. The current growth rate of the catalogue
is around 50 entries per month, mostly from the CVcat editors.

The data flow into CVcat originating from outside the CVcat core
team is not yet satisfying. It is unclear why most users refrain from adding
data from their own publications. Users should bear in mind that the
probability of having their papers cited increases if the information from
their publications can be found in the database.

\section{Prospects}
\label{s:prospects}

The concept of CVcat has demonstrated the benefits of a public scientific
catalogue with a globally-shared expertise of its users and editors. The
general structure of CVcat is also applicable to other fields of
astronomy. Hence, a more general software based on the experiences with CVcat
which will allow the implemenation of catalogues for arbitrary object classes,
called ``AstroCat'', is currently being developed. This software will also include
additional features to improve the knowledge management of astronomical
results and data:
\begin{itemize}
\item Besides the ``single number data'' stored so far, the infrastructure for
storing more complex data products, e.g., light curves, spectra, and finding
charts will be added. This feature should help to overcome a major shortcoming
in the present method of scientific publishing: most authors publish their
reduced data only in the form of plots, making follow-up work on these results
rather unattractive. While these data \emph{may} be obtained directly from the
authors, experience shows that in many cases the data have been lost due to
faulty hardware or storage media, with the probability of loss increasing
dramatically with time since the publication of the original work. In
addition, the authors may have left astronomy, preventing access to their
data.  If the data is available in CVcat, the re-examination of observations
would be much easier and independent of contact with the original
author and the corresponding data archive. The storage of reduced and published data in a
usable form at e.g. CDS, Strasbourg, is already promoted by journals like A\&A.
\item Information agents will allow users to be informed automatically if new
data is entered into the database for their objects of interest. This is a
service that keeps the user informed about new publications on a given
object or class of objects without having to log in to CVcat frequently.
\item An elaborated validation system will allow the users to add their
comments to published data. This unique feature will add
personal communication aspects to the database. Public discussion on details of the
published data may arise from this.

\item
To allow exact quoting of a specific state of the ever-changing
catalogue, a ``versioning´´ method using a global identification number (GID) will be used: the
GID is incremented with each change of the content of AstroCat (e.g. adding new data, marking data as correct, marking data as the best available etc.). Every output resulting from the use of AstroCat will include
the current GID number. An older state of the database can be exactly
reproduced by issuing a given GID (lower than the current one), and
using AstroCat with such a specific GID will return always
\textit{precisely} the same results, independent of the actual state
of the data base. Any statistical analysis based on data from CVcat
should therefore include the GID of the CVcat state at the time of analysis, permitting a quantitative comparison with future studies. It is not necessary to rely on a fixed or frozen state so far only available in printed
catalogues.

\item Data that have not yet been cross-checked by the editorial team will be
tagged as such. This method combines both the unique speed of the CVcat
concept and the expertise of the editorial team. As long as a new datum is
visible but marked as unapproved, the user of the database can use the new
entry under his own responsibility, while an approved value is authoritative.
\item Hyperlinks to a large number of available web resources will be included
for every object. This includes, e.g., links to other CV cataloges like SIMBAD, the ADS, arXiv.org, the different
variable star observers archives, and many more.
\end{itemize}

The development of the new CVcat and the AstroCat software is done in close
cooperation with the users of the current implementation. In summer 2003, the
data from the current CVcat will be transferred to the new database. By the
end of 2003, the AstroCat/CVcat framework will be completed. Ideas from the
editors and users of other interactive catalogues, e.g. the high-$z$ database
\citep{HighZ}, are highly appreciated. The AstroCat project is hosted at
\texttt{http://astrocat.uni-goettingen.de}.

\section{Technical realization}
CVcat is implemented as a Perl script which runs on a Linux PC. This Perl
script processes the user requests (HTTP GET/POST requests) and creates HTML
pages that are delivered via the Apache web server. Hence, CVcat presents
itself as an interactive web page.

The data is stored in a MySQL data base to which the Perl script communicates
using the standard Perl DBI/DBD interface.

For the implementation of AstroCat, an XML layer will be included between the
data base and the HTML layer. This XML layer can be used to automatically
include larger data sets into the database. Another possible application will
be to use the AstroCat framework as an archive for reduced data from robotic
telescopes that use RTML \citep{RTML} for their observation requests,
accomplished observations, and some of the metadata. A technical advantage of
the usage of XML is the easy availability of many very good tools specialized
in the processing of XML data and the transformation of XML documents into
HTML pages. This will improve the quality and the speed of the
implementation. PHP might be used in addition to Perl as the scripting
language for the AstroCat programs.

\section{Summary}

With CVcat we have implemented an online catalogue for cataclysmic variable
stars, which -- for the first time -- allows its users to add data instantly
visible to all other users. The quality control is realized by a team of
experts who share the responsibility for the different object classes. From
the experiences of one year of public usage of the catalogue we have compiled
a number of new concepts which will be implemented in the next generation of
CVcat in the context of a more general framework for astronomical databases,
AstroCat.

\begin{acknowledgements}
We would like to thank all the colleagues who volunteered to take the
responsibility for one of the object classes included in CVcat: Domitilla de
Martino, Don Hoard, Steve Howell, Tom Marsh, Klaus Reinsch, Matthias
Schreiber, and John Thorstensen. The AstroCat/CVcat project is funded by the
\emph{Deut\-sche For\-schungs\-ge\-mein\-schaft (DFG)\/} project number
LIS~4~--~554~95~(1)~G\"ottingen. The development of the feasibility study has
partially been supported by the DLR under grant 50\,OR99\,03\,6. BTG also
acknowledges support from a PPARC Advanced Fellowship.

We appreciate the helpful comments from the referee, Dr. Ochsenbein, to both
this paper and the CVcat database itself.
\end{acknowledgements}

\bibliographystyle{3352}

\end{document}